\documentclass[runningheads]{llncs}

\usepackage{graphicx}

\usepackage{hyperref}
\usepackage{breakurl}

\usepackage{arydshln}
\usepackage{multirow}
\usepackage{wrapfig}
\usepackage{longtable}
\usepackage[table,xcdraw]{xcolor}
\usepackage{float}

\begin{document}

\title{Relational Expressions for Data Transformation and Computation}

\author{David Robert Pratten\inst{1}\orcidID{0000-0001-9210-9529} \and Luke Mathieson\inst{1}\orcidID{0000-0001-6470-2296}  }

\authorrunning{D. R. Pratten and L. Mathieson}

\institute{University of Technology, Ultimo NSW 2007, Australia
\email{david.r.pratten@student.uts.edu.au}\\
\email{luke.mathieson@uts.edu.au}\\
}

\maketitle

\begin{abstract}

\sloppypar
Separate programming models for data transformation (declarative) and computation (procedural) impact programmer ergonomics, code reusability and database efficiency. To eliminate the the necessity for two models or paradigms, we propose a small but high-leverage innovation, the introduction of complete relations into the relational database. Complete relations and the discipline of constraint programming, which concerns them, are founded on the same algebra as relational databases.
We claim that by synthesising the relational database of Codd and Date, with the results of the constraint programming community, the relational model holistically offers programmers a single declarative paradigm for both data transformation and computation, reusable code with computations that are indifferent to what is input and what is output, and efficient applications with the query engine optimising and parallelising all levels of data transformation and computation.

\end{abstract}

\section{Introduction}

Separate programming models for data transformation and computation impact programmer ergonomics, code reusability and database efficiency.

Concerning programmer ergonomics, when building business applications, programmers frequently switch between declarative SQL\footnote{SQL was chosen as an illustrative vehicle rather than Datalog, or some other relation expression language, due to its ubiquity.} code to transform data and then to a procedural programming language for the computation that encodes business logic and rules. Programmers' daily work is switching back and forth between these two paradigms.

With regard to reusability, data in a relational database sets a high bar not reached by procedural code. Once we have captured data in a set of relations, we have high confidence that we can use queries to answer arbitrary questions in the future. Procedural code does not have this ability.
To illustrate this point, if we had procedural code that computed property sale stamp duty for the Australian Capital Territory (ACT) based on the sale price and multiple factors, we would be unlikely to be able to use that same code to compute the reverse query and tell us the possible sale prices if we knew the duty paid.

Finally, database procedural code imposes a performance penalty because query engines cannot introspect and optimise it. This issue is being addressed (after the fact) by efforts such as~\cite{Emani2016ExtractingApplications}, \cite{Hirn2020PL/SQLPL}, and~\cite{Zhang2023AutomatedSQL}, which translate procedural code back into relational algebraic expressions including fixed-point operators.

To eliminate the the necessity for two models or paradigms, we propose a small but high-leverage innovation, the introduction of complete relations into the relational database.  A relation is complete if it has the value formed as the cross-product of all its attribute's domains.\footnote{The related term ``complete database'' refers to a relational database where tuples are restricted to be drawn from the corresponding complete relation, i.e. NULL values are prohibited.~\cite{Libkin2016SQLsAnswers}} Complete relations and the discipline of constraint programming, which concerns them, are founded on the same algebra as relational databases. Using a biological metaphor, complete relations may be likened to a stem cell, which, when coupled with a suitable predicate, may be specialised to become any computable relation.  We introduce the term ``sigma complete relation'' to denote complete relations that have been specialised by the relational \(\sigma\) operator to capture a specific set of business rules and logic.

In this paper, we review related work that has previously explored the intersection of relational databases and constraint programming. (Section \ref{sec:RelatedWork}) After demonstrating the relational database's and constraint programming's common foundations and crystalising the challenge that we are addressing (Section \ref{sec:PreliminaryandProblemDefinition}), we explore the core technical innovation of this paper, the sigma complete database relation. (Section \ref{sec:SigmaCompleteRelations}).  We reproduce the results of a recent proof of concept by Salsa Digital for the OECD's Observatory of Public Sector Innovation using sigma complete relations, demonstrating single paradigm programming and the flexibility of querying sigma complete relations. (Section \ref{sec:ExperimentalStudy})

\section{Related Work} \label{sec:RelatedWork}
Constraint programming emerged as a generalisation of ``inferential'' and ``deductive'' approaches explored by the logic programming community\cite{Jaffar1987ConstraintProgramming}. % DONE Reference needed
In contrast, Codd carefully excluded such ``inferential'' and ``deductive'' systems from discussion in his seminal paper~\cite{Codd1970ABanks} on relational databases. Despite this separation at the foundation, in the relational database and constraint programming corpora, we find discussion of the connection of database relations with constraints and computation under three rubrics: Algorithmic Relations, Computed Relations, and Constraint Database.
These prior works appear to be pushing toward the same intuition that underlies this paper.

\subsubsection{Algorithmic Relations} \label {Algorithmic Relations}
In 1975 Hall, Hitchcock and Todd~\cite{Hall1975AnComputation} introduced algorithmically represented relations as a complement to relations stored as a set of tuples. One of Hall, Hitchcock and Todd's motivations was the power of algorithmic relations to compute multi-directional queries being indifferent as to which attributes are input and which are output. They contributed the notion of a relation's ``effectiveness''. Effectiveness refers to the subset of grounded attributes for which the relation will define a complete set of tuples. For data relations, the effectiveness is always the powerset of attributes, meaning that if we ground none, any, or all attributes, then the data relation will yield all relevant tuples. For algorithmic relations, this is not assured. Hall, Hitchcock and Todd classified the subject of this paper as ``pure predicate relations'', however they appear unaware that these relations contain enough information for an evaluation engine to generate their extensions. No record of an implementation or further exploration of algorithmic relations has been found in the literature.

\subsubsection{Computed Relations} \label {Computed Relations}
Across the Atlantic, computed relations were separately described by David Maier and David Warren in 1981~\cite{Maier1981IncorporatingDatabases} as a way of extending the standard relational theory of databases with more computation power. Maier and Warren motivated their proposal by considering the challenges of including arithmetic in the relational algebra and also use cases for posing reverse queries. Maier and Warren introduced the definitions and theorems to incorporate computed relations into the relational model, and their theory and algorithms fore-sage the comprehensive treatment that would subsequently be developed independently within the discipline of constraint programming. Maier and Warren showed how selection predicates and join conditions are vital context that enables a query engine to yield a listable (finite) extension from a computed relation. No record of an implementation or further exploration of computed relations has been found in the literature.

\subsubsection{Constraint Database CD} The logic and constraint programming communities recognise the equivalence of first-order\footnote{Constraint programming constructs outside of first-order logic are not considered in this paper as Relational Algebra is equivalent to Relational Calculus which can be expressed as formulas in first-order logic. \cite[p241]{Garcia-Molina2009DatabaseBook} %Ref for this too if possible
} constraint programming with relational algebra, see for example~\cite{Kanellakis1994ConstraintLanguages} and~\cite{Kolaitis2003ConstraintLogic}.  Bussche, one of the contributors to~\cite[p35]{2000ConstraintDatabases} after consideration of related evaluation mechanisms, concluded that ``the relational algebra serves as an effective constraint query language''.
Cai in~\cite[p173]{Gai2004IntegratingSystems} discussed how to incorporate constraints within a relational database and reiterated the point made by Maier and Warren that it is vital for context to be shared from the relational expression with the constraint evaluation engine.

The practical outworking of this recognition has been in a model called ``Constraint Database''. Quoting Kuper, Libkin, and Paredaens~\cite[p7]{2000ConstraintDatabases} ``the basic idea of the constraint database model is to generalize the notion of a tuple in a relational database to a conjunction of constraints''.  In this model, linear or polynomial equations represent large or infinite sets compactly, which matches well to spatiotemporal applications~\cite{Revesz2010IntroductionSpatio-Temporal}.

While sigma complete relations and CD constraint relations are clearly adjacent concepts, after reviewing Revesz's 2010 survey of the model~\cite{Revesz2010IntroductionSpatio-Temporal}, Table~\ref{tab:compareSCRwithCDrelation} summarises significant differences between them.
\begin{table}[htp]
\centering
\begin{tabular}{|p{6cm}|p{6cm}|}\hline
 \textbf{A sigma complete relation} & \textbf{ A CD constraint relation}                                   \\ \hline
The relation is central to unifying the relational expression with constraint programming & The relation is aimed at compact representation and querying of segments and low order volumes \\ \hline
The focus is on complete relations, which may be infinite. & The focus is on values that consist of infinite points. \\ \hline
The complete relation may be constrained, as a whole relation, by a Boolean expression.                 & Each tuple may contain a different constraint. \\ \hline
The complete relation can't hold any data belonging to any entity in the database.                      & Each tuple may hold user data in addition to a constraint.             \\ \hline
The relation is a constant value.                                                              & The relation is subject to normal DML operations such as \verb|INSERT|, \verb|DELETE| and \verb|UPDATE|. \\\hline
\end{tabular}
\caption{Contrast of sigma complete relations with CD's constraint relations.}
\label{tab:compareSCRwithCDrelation}
\end{table}

According to the Scopus database, publications on research into Constraint Databases, including proof of concept implementations, were most frequent for a decade from the mid-1990s. See \cite{Revesz2010IntroductionSpatio-Temporal} for a survey of this work.

\section{Preliminaries and Problem Definition} \label{sec:PreliminaryandProblemDefinition}
The relational database and constraint programming communities' relative isolation is a historically contingent fact that belies their common foundation in the relational model.
At the most fundamental level, the difference between the two communities is that one considers an empty data relation the starting point for work. In contrast, the other considers the complete relation the place to begin. The characteristic challenges being solved by each discipline have emerged naturally from their different starting points.

\subsection{Database Relational Model}  Along with data relations, today's relational databases have already incorporated aspects of computation. Following Date~\cite{Date2019DatabaseJazz}, the relational database model views a relation as a set of n typed attributes (constituting the relation’s heading), along with a set of n-ary tuples (constituting the body).

In addition to a heading and a body, each data relation also has a predicate which represents the intended interpretation of the relation. The predicate is a natural language statement always true for the relation. An example of a predicate might be ``Is one of our organisation's customers'' for a relation called \verb|cust|. Data relation values are empty until tuples are added by a relational expression or a Data Manipulation Language (DML) statement such as \verb|INSERT|.

\sloppypar
For partially enforcing a data relation's predicate, the SQL standard provides four integrity constraint types: \verb|PRIMARY KEY|, \verb|UNIQUE|, \verb|CHECK|, and \verb|FOREIGN KEY|~\cite{Kline2022SQLNutshell}.
For expressing computation, SQL adds features\footnote{Fixed-point operators\cite{Eisenberg1999SQL:1999SQL3} and string operators are not considered in the current work and are open questions for further investigation.} beyond the relational algebra, including arithmetic expressions, derived attributes  (a.k.a. generated columns)~\cite{Eisenberg2004SQL:2003Published}, table-valued functions, and stored procedures~\cite{Gupta2021ProceduralWild}.

\subsection{Constraint Programming} The Constraint Programming community has also adopted the relation as an abstraction; however, their starting place is the complete relation. Quoting Hooker's survey\cite[p. 376]{Hooker2012IntegratedOptimization}, the parallels are immediate: ``... a constraint can be viewed as a relation, i.e., a set of tuples belonging to the Cartesian product of the variable domains''.

Following Marriott and Stuckey~\cite{Marriott1998ProgrammingIntroduction}, Frisch and Stuckey~\cite{Frisch2009TheLanguages}, Stuckey and Tuck~\cite{Stuckey2013MiniZincFunctions}, and as embodied in the \href{https://www.minizinc.org/}{MiniZinc} language, constraint programming views constraint relations as a set of arguments, each with their domain (type) and with a Boolean expression as a constraint over these arguments. When we ask, ``What satisfies this constraint relation?'' the answer is the constraint relation's extension, a set of solutions with zero, finite, or infinite cardinality. This foundation for constraint programming is sufficient to enable it to express arbitrary\footnote{Computations within the scope of the logical theories\cite{Barrett2018SatisfiabilityTheories} embodied in the evaluation engine.} computations in a declarative fashion. In contrast to data relations, constraint relations are always constant values.

For the benefit of relational database practitioners, who may not be familiar with constraint programming, here are two examples to illustrate these concepts. The first annotated example is written in the MiniZinc language~\cite{Stuckey2023TheHandbook}. It implements a business rule that says that all Australian states and territories should have different (one of three) colours when compared with each adjacent state.  The code is based on the example in section 2.1.1 of the MiniZinc Handbook[ibid].

\scriptsize \begin{verbatim}
    int: number_of_colours = 3;
    predicate colour_Australia(
        var 1..number_of_colours: wa, var 1..number_of_colours: nt,
        var 1..number_of_colours: sa, var 1..number_of_colours: qld, 
        var 1..number_of_colours: nsw, var 1..number_of_colours: vic)
    = let {
        constraint wa != nt;
        constraint wa != sa;
        constraint nt != sa;
        constraint nt != qld;  constraint sa != qld; constraint sa != nsw;
        constraint act != nsw; constraint sa != vic; constraint qld != nsw;
        constraint nsw != vic
    } in true;
\end{verbatim} \normalsize

\begin{table}[htp]
\centering
\begin{tabular}{|l|l|l|l|l|l|l|}
\cline{1-7}
\textbf{wa} & \textbf{nt} & \textbf{sa} & \textbf{qld} & \textbf{nsw} & \textbf{act} & \textbf{vic} \\ \cline{1-7}
3 & 2 & 1 & 3 & 2 & 1 & 3 \\ \cline{1-7}
3 & 2 & 1 & 3 & 2 & 3 & 3 \\ \cline{1-7}
2 & 3 & 1 & 2 & 3 & 1 & 2 \\ \cline{1-7}
2 & 3 & 1 & 2 & 3 & 2 & 2 \\ \cline{1-7}
3 & 1 & 2 & 3 & 1 & 2 & 3 \\ \cline{1-7}
3 & 1 & 2 & 3 & 1 & 3 & 3 \\ \cline{1-7}
1 & 3 & 2 & 1 & 3 & 1 & 1 \\ \cline{1-7}
1 & 3 & 2 & 1 & 3 & 2 & 1 \\ \cline{1-7}
2 & 1 & 3 & 2 & 1 & 2 & 2 \\ \cline{1-7}
2 & 1 & 3 & 2 & 1 & 3 & 2 \\ \cline{1-7}
1 & 2 & 3 & 1 & 2 & 1 & 1 \\ \cline{1-7}
1 & 2 & 3 & 1 & 2 & 3 & 1 \\ \cline{1-7}
\end{tabular}
\caption{Colouring Australian states and territories in three colours.}
\label{tab:colouraus}
\end{table}

When asked to supply all solutions, MiniZinc will respond with the
twelve solutions shown in Table \ref{tab:colouraus}. Note that if we omitted the \verb|constraint|s, the extension of the constraint relation would have been the cross product of the attribute's domains i.e. the complete relation.

The second example implements the business rule for calculating Australian Goods and Services Tax (GST). The annotated \verb|Australian_GST| constraint relation is of arity three. It captures the relationship between consumer \verb|Price|, the Goods and Services Tax \verb|GSTAmount|, and the price before applying the GST \verb|ExGSTAmount|. This constraint relation is also written here in MiniZinc:
\scriptsize \begin{verbatim}

predicate Australian_GST(
    var float: Price,
    var float: ExGSTAmount,
    var float: GSTAmount) =
let {
    constraint Price/11 = GSTAmount;
    constraint ExGSTAmount = Price-GSTAmount;
    }
in true;

\end{verbatim} \normalsize
Note that even with the equation-like constraints in place, the \verb|Australian_GST| constraint relation, while no longer complete, is of infinite cardinality.

\subsection{The Database Relational Model and Constraint Programming}
In this section, we will begin by clarifying terminology and then show their common foundations. In table \ref{tab:cognates}, cognate terms from the database relational model and those from constraint programming are shown along with the terms that will be used in the rest of this paper.  Alternate terms found in the literature are included in italics.

\begin{table}[htp]
\centering
\begin{tabular}{|p{2cm}|p{5cm}|p{5cm}|}
\hline
\textbf{Term} & \textbf{Database Relational Model Cognates} & \textbf{Constraint Programming Cognates} \\ \hline
\textbf{Relation} & Relation \textit{or Table} & Predicate \textit{or Constraint Relation}  \\ \hline
\textbf{Predicates \hfill \newline   \hfill \newline \hfill \newline } & Predicate (natural language) \hfill \newline \hfill \newline Integrity constraints (Boolean expressions) &  \hfill \newline \hfill \newline Constraints (Boolean expressions)\\ \hline
\textbf{Heading} & Heading \textit{or Columns or Arguments} & Arguments, Variables or Parameters \\ \hline
\textbf{Argument} & A single Argument \textit{or Column, Generated Column} & A single Argument, Variable or Parameter \\ \hline
\textbf{Domain} & Type \textit{or Domain} of an argument& Domain \textit{or Type} of an argument \\ \hline
\textbf{Extension}&Body \textit{or Tuples or Rows} & Solutions \textit{or Extension} \\ \hline
\textbf{Tuple} & Tuple \textit{or Row} & Solution \\ \hline
\end{tabular}
\caption{The relational model terminology used in the rest of this paper.}
\label{tab:cognates}

\end{table}

After comparing the literature referenced above and after discounting purely linguistic distinctions, the commonalities, and some unique features, between the database relational model and constraint programming are summarised here:
\begin{itemize}
    \item Both are based on the mathematical relation
    \item Their relations have a header with attributes, each with defined domains
    \item Their relations have predicates that describe the meaning of the relation. (For data relations, this is typically a natural language condition, whereas, for constraint programming, the predicate is explicit and computable.)
    \item Their relations may be defined by their extensions. In SQL, using DML statements such as \verb|INSERT|, \verb|DELETE| and \verb|UPDATE|. In MiniZinc's case, by providing a data file~\cite[2.1.3]{Stuckey2023TheHandbook}.
    \item Both may be constants, however, constraint programming relations are always constant values.
    \item Both expression languages obey the relational algebra, including the familiar relational operators: \(\pi\) project, \(\sigma\) select, \(\times\) cross-product, and the compositions of these operations such as \(\bowtie\) join
    \item Both approaches are declarative, relying on an evaluation engine to take the declarative expression and construct an efficient evaluation strategy based on relational algebraic properties and heuristics. In MiniZinc's case, after applying algebraic transformations, solutions are found using a variety of back-end solvers. For example: \href{https://www.gecode.org}{Gecode},  \href{https://github.com/chuffed/chuffed}{Chuffed}, \href{https://developers.google.com/optimization/}{OR-Tools}.
\end{itemize}

\subsection{[Re]Uniting Two Worlds}

We now finally arrive at a point where we can state the central problem and central thesis of this paper. Programmers have reaped only half the available benefits from the relational model. By adopting the relational database of Date and Codd, programmers gained independence\cite{Codd1970ABanks} from providing a procedural access path to the data of interest and have been able to specify data transformations in a declarative fashion. The other potential benefit of the relational model is to free programmers from having to code computations and business rules procedurally, and this benefit remains to be realised.

We claim that by synthesising the relational database of Codd and Date, with the results of the constraint programming community, the relational model holistically offers programmers a \textit{single declarative paradigm} for both data transformation and computation, \textit{reusable code} with computations that are indifferent to what is input and what is output, and \textit{efficient applications} with the query engine optimising and parallelising \textit{all levels} of data transformation and computation.

\section{Sigma Complete Relations} \label{sec:SigmaCompleteRelations}
\subsubsection{Formal Definition} Formally, a sigma complete relation \verb|S| may be defined using relational algebra as:

\[ S = \sigma P (Dom(a_1) \times Dom(a_2) \times ... \times Dom(a_N)) \]
for predicate \verb|P| and the domains \(Dom()\) of the attributes \({a_1, a_2, ... a_N}\). Depending on the domains of attributes, the sigma complete relation may have a finite or infinite cardinality.

\subsubsection{Illustrative Examples} Let's re-express the MiniZinc \verb|Australian_GST| example to illustrate this. Firstly we look at the example through the idiom of SQL and, secondly, through the Tutorial-D\cite{Date2014TheManifesto} language of Date and Darwen. We may begin by letting the following form represent a complete relation with an arity of three:
\scriptsize \begin{verbatim}
COMPLETE(Price float, ExGSTAmount float, GSTAmount float)
\end{verbatim} \normalsize
As it stands, the \scriptsize\verb|COMPLETE(Price float, ExGSTAmount float, GSTAmount float)|\normalsize relation is infinite, so the following relation expression would \emph{not} be listable.
\scriptsize \begin{verbatim}
SELECT * FROM COMPLETE(ExGSTAmount float, GSTAmount float, Price float);
\end{verbatim} \normalsize

We can now express the Australian GST rule as a specialisation of this complete relation. The required predicate is placed in the \verb|WHERE| clause, noting, of course, that this is a Boolean expression, not a pair of assignment statements.
\scriptsize \begin{verbatim}
SELECT *
    FROM COMPLETE(Price float, ExGSTAmount float, GSTAmount float)
    WHERE GSTAmount = Price/11 AND ExGSTAmount = Price-GSTAmount;
\end{verbatim} \normalsize
If we supply further information by grounding (fixing) one of its attributes, the expression gives a listable extension.  Here is an example GST calculation for \verb|Price = 110|:
\scriptsize \begin{verbatim}
SELECT *
    FROM COMPLETE(Price float, ExGSTAmount float, GSTAmount float)
    WHERE GSTAmount = Price/11 AND ExGSTAmount = Price-GSTAmount
        AND Price = 110;
\end{verbatim} \normalsize
The single tuple result is shown in Table~\ref{tab:my-table8}.
\begin{table}[htp]
\centering
	\begin{tabular}{|l|l|l|}
		Price & ExGSTAmount & GSTAmount \\ \hline
		110   & 100         & 10        \\
		\hline
	\end{tabular}
	\caption{Australian\_GST relation expression result}
 \label{tab:my-table8}
\end{table}
For convenience and reuse, we can give the relation expression the name \verb|Australian_GST| within our database:
\scriptsize \begin{verbatim}
CREATE VIEW Australian_GST AS
    SELECT *
        FROM COMPLETE((Price float, ExGSTAmount float, GSTAmount float)
        WHERE GSTAmount = Price/11 AND ExGSTAmount = Price-GSTAmount;
\end{verbatim} \normalsize

When evaluated, any of the following relation expressions will give the same result as above!
\scriptsize \begin{verbatim}
select * from Australian_GST where ExGSTAmount = 100;
select * from Australian_GST where GSTAmount = 10;
select * from Australian_GST where Price=110;
\end{verbatim} \normalsize

While less familiar to relational database practitioners, the constraint programming community will recognise the above as a straightforward application of satisfaction solving. The relation expression engine is \textbf{not} retrieving all rows of the infinite \verb|Australian_GST| relation and then selecting the required tuple. Instead, predicate, or selection, push-down\cite{Ullman1988PrinciplesI} is being used to strengthen the constraints on the sigma complete relation sufficiently to yield the required answer without search.

In recent\footnote{by email 25 July 2023} private correspondence on the topic of this article with C.J. Date and Hugh Darwen, they offered the following extension to their published Tutorial D syntax as a way of incorporating sigma complete relations in the database:
\scriptsize \begin{verbatim}
RELATION {Price float, ExGSTAmount float, GSTAmount float}
    WHERE GSTAmount = Price/11 AND ExGSTAmount = Price-GSTAmount
\end{verbatim} \normalsize
They noted that within Tutorial D, the sigma complete relation may appear as a database component by
declaring it as a name constant, prefixing the above expression by \verb|CONST AustralianGST|, giving:
\scriptsize \begin{verbatim}
CONST AustralianGST
    RELATION {Price float, ExGSTAmount float, GSTAmount float}
        WHERE GSTAmount = Price/11 AND ExGSTAmount = Price-GSTAmount
\end{verbatim} \normalsize

There are further worked examples available on GitHub including an  \href{https://github.com/DavidPratten/jetisu/blob/main/ACT_Conveyance_Duty.ipynb}{Australian Capital Territory (ACT) Conveyance Duty} sigma complete relation which encodes the multi-tiered and multi-variable contingent relationship between price and conveyancing duty in the Australian Capital Territory from July 1st, 2022. The single relation is equally able to answer the question ``How much duty is chargeable for a property valued at \$1.2M?'' and its inverse, ``A property owner pays \$140,740 in conveyancing duty. What were the possible property prices?''

\subsubsection {Implementation Considerations} To our knowledge, no one has yet built a holistic relation expression evaluation engine that operates over both data relations and complete relations. However, to use a botanical metaphor, building such an engine is likely to be akin to the grafting together of two stems rather than requiring radical replanting. The following observations support this:

\begin{itemize}
    \item Sigma complete relations are subject to the same algebraic transformations already used in relational database engines. For example, sigma complete relations may be projected over a subset of their arguments.
    \item Optimisation techniques over data relations and complete relations are closely related. e.g. constraint propagation \cite{Bessiere2006ConstraintPropagation} is closely related, and may be identical to, the relational database optimiser's predicate push-down/migration strategy.
    \item Sigma complete relations have a well-defined \(\bowtie\) semantic close to, and possibly identical to, SQL:1999's \verb|LATERAL|.
\end{itemize}

It is envisaged that the heuristics that guide expression simplification and optimisation will be where most of the effort will be.  The first step in the implementation work will be to identify a harmonious minimum set of theories, domains, algorithms, and heuristics from relational databases and constraint programming languages.

\subsubsection{Termination Considerations}
Sigma complete relations may be of infinite cardinality, and this poses questions about the termination of the evaluation of relational expressions. In practice, it is found that the set of useful relational expressions over sigma complete relations correlates with expressions that provide just enough grounding in the expressions to enable the query engine to provide a useful (finite) extension. However, if a relation expression insufficiently constrains a sigma complete relation of infinite extent, its evaluation behaviour would be similar to SQL:1999, where a query will not terminate if a recursive common table expression fails to find a fixed point.

Due to the unavailability of a holistic relational expression evaluation engine, all the experiments in this paper have evaluated SQL relational queries over sigma complete relations defined by predicates in MiniZinc. They therefore are unable to fully test the single programming environment thesis of this paper. However, let's see what we can learn from evaluating this early prototype.

\section{Experimental Study} \label {sec:ExperimentalStudy}
This evaluation replicates the results of an internationally recognised proof of concept (PoC) and passes a comprehensive test suite with some advantages over the original implementation. An open-source project hosts all the code and instructions for repeating the evaluation using Docker.

Late in 2022, the OECD's Observatory of Public Sector Innovation published Salsa Digital's PoC``Delivering a personalized citizen experience using Rules as Code as a shared utility''~\cite{ObservatoryofPublicSectorInnovationDeliveringUtility}. The site ``Interactive Q\&A with COVID Rules for Workplaces'' was coded using OpenFisca~\cite{AgenceNationaledelaCohesiondesTerritoires2022OpenFiscaCode},~\cite{SalsaDigitalWhatOpenFisca}. Salsa Digital host an interactive version of the PoC on their website.~\cite{SalsaDigitalCoronavirusConcept}.

Sigma complete relations were evaluated using the \href{https://github.com/DavidPratten/jetisu}{Jetisu Toolkit} by David Pratten. While not ideal, to facilitate fast prototyping, this implementation used MiniZinc to code the sigma complete relation with its Boolean predicate. SQL is used for querying the relation.

\textbf{Am I up to date?} The evaluation modelled rules for up-to-dateness of COVID vaccinations as a relation:

\scriptsize \begin{verbatim}
CREATE VIEW covid_vaccinations AS
    SELECT * FROM COMPLETE(age int,
                recommended_doses int,
                booster_doses int,
                vaccine_doses intr,
                immuno_compromised_disability bool,
                months_since_last_dose_at_least int,
                eligibility bool,
                up_to_date bool
                )
        WHERE <Boolean expression>;
\end{verbatim} \normalsize
The definition of the relation is 36 lines of MiniZinc code and may be found at github.com \href{https://github.com/DavidPratten/jetisu/blob/main/jetisu/covid_vaccinations.mzn}{code link}. Multiple relational expressions and their results may be found on this page: \href{https://github.com/DavidPratten/jetisu/blob/main/COVID_vaccinations.ipynb}{Australian COVID Vaccination eligibility and up-to-date (circa mid-2022)}.

\textbf{Do I need COVID-19 vaccination for my workplace?}
The evaluation captured workplace-related rules across three Australian state jurisdictions: New South Wales, Victoria, and Western Australia with this relation:
\scriptsize \begin{verbatim}
CREATE VIEW covid_vaccinations_and_work AS
    SELECT * FROM COMPLETE(work_location work_location_enum,
                    work_sector work_sector_enum,
                    specialist_school bool,
                    aged_care_facility bool,
                    disability_worker_in_school bool,
                    private_home_only bool,
                    nsw_health_worker bool,
                    mandatory bool,
                    doses int
                    )
        WHERE <Boolean expression>;
\end{verbatim} \normalsize

The definition of the relation is 99 lines of MiniZinc code and may be found at github.com \href{https://github.com/DavidPratten/jetisu/blob/main/jetisu/covid_vaccinations_and_work.mzn}{code link}. The page \href{https://github.com/DavidPratten/jetisu/blob/main/COVID_vaccinations_and_work.ipynb}{Australian COVID Vaccinations and Work (circa mid-2022)} shows how relation expressions may be used to ask inverse or abductive questions such as ``What are the vaccination requirements for emergency service workers across the three jurisdictions?''

\textbf{Interactive Q\&A} As referenced above, the original PoC by Salsa Digital makes their Rules as Code model available to the public as a Question and Answer dialog. The \href{https://github.com/DavidPratten/jetisu/blob/main/Goal_seeking_covid_vaccination_and_work.ipynb}{Interactive Q\&A with COVID Rules for Workplaces} page shows how sigma complete relations support this use-case. The evaluation goes one step ahead of Salsa Digital's PoC here.  The availability of the ruleset as a relation enables the code to automatically minimise the number of questions asked by ordering the questions based on which question, if asked next, will narrow the number of alternatives the fastest.

\textbf{Evaluation Summary}
The evaluation demonstrated that sigma complete relations can replicate PoC results created with a state-of-the-art tool, OpenFisca.  The sigma complete relation passes all 103 cases of the PoC's test suite, replicates the Q\&A ability and has the following advantages over the solution with OpenFisca:
\begin{itemize}
    \item Automatic question sequencing based on the fastest path to an answer
    \item After discounting for OpenFisca's presentation level code, the code size for the sigma complete relations evaluation was an order of magnitude less than for OpenFisca. (36 vs 293 lines for \verb|covid_vaccinations|, and 99 vs 520 lines for \verb|covid_vaccinations_and_work|)
    \item Supports querying in the forward and reverse (abductive) directions.
\end{itemize}

\section{Conclusion} \label {Open Questions}

This paper is an early report on using relational expressions for data transformation and computation, inviting feedback from researchers in both the relational database and constraint programming communities. We are in a line of researchers reaching for an intuition that is not yet fully revealed, and much remains to be explored and developed. As you have seen, this paper is an invitation to work towards a programmer experience where a declarative approach is used for all aspects of data transformation and computation.  Where code is written once and used for many, as yet, un-thought-of queries. And where the execution engine has visibility of the whole computation stack and can optimise from top to bottom. The evaluation reported above suggests that this is a path worth following.

There are many questions surrounding sigma complete relations that recommend themselves for further study. One issue is to identify a minimum set of solving theories\cite{Barrett2018SatisfiabilityTheories} that give programmers a good experience when utilising complete relations in the database. Research into relational algebraic languages is a vibrant field. New alternatives to SQL now include languages such as  \href{https://learn.microsoft.com/en-us/dotnet/csharp/linq/}{LINQ}, \href{https://prql-lang.org/}{PRQL}, \href{https://github.com/malloydata/malloy}{Malloy}, \href{https://github.com/julianhyde/morel}{Morel}, \href{https://www.edgedb.com/}{EdgeDB}, and \href{https://learn.microsoft.com/en-us/powerquery-m/}{Power Query's M formula language}. A related question is to evaluate which relational expression language would be most approachable for an initial database implementation of sigma complete relations.  The relational model is associated with concepts like normalisation, functional dependencies, candidate keys and primary keys. A question related to these concepts is how sigma complete relations increase the power of relational data modelling. And finally, what is the best way to extend the reported initial first-order results to cover aggregations (SQL \verb|GROUP BY|) and predicates over aggregations (SQL \verb|HAVING|) and then higher-order optimisation and combinatorial problems \cite{Mancini2012CombinatorialDatabases} and \cite{Brucato2016ScalableSystems}?

\subsubsection*{Acknowledgements} \label {Acknowledgements}
Special thanks to Hugh Darwen without his encouragement and sharp insights, this paper would not have seen the light of day. And to the team behind MiniZinc for such a beautiful tool. Thanks to the management and staff of Salsa Digital for access to the Python code base behind their OECD Case Study.
\newpage

\bibliographystyle{splncs04}
\bibliography{references}

\end{document}